\documentclass{vgtc}                          

\ifpdf
  \pdfoutput=1\relax                   
  \usepackage{graphicx}                
  \DeclareGraphicsExtensions{.pdf,.png,.jpg,.jpeg} 
\else
  \usepackage{graphicx}                
  \DeclareGraphicsExtensions{.eps}     
\fi%

\graphicspath{{figures/}{pictures/}{images/}{./}} 

\usepackage{microtype}                 
\PassOptionsToPackage{warn}{textcomp}  
\usepackage{textcomp}                  
\usepackage{mathptmx}                  
\usepackage{times}                     
\usepackage{cite}                      
\usepackage{tabu}                      
\usepackage{booktabs}                  

\onlineid{1108}

\vgtccategory{Research}

\vgtcinsertpkg

\title{\datatales: Investigating the use of  Large Language Models \\ for Authoring Data-Driven Articles}

\author{
    Nicole Sultanum\thanks{email: nsultanum@tableau.com}  
\and
    Arjun Srinivasan\thanks{email: arjunsrinivasan@tableau.com}  
}
\affiliation{\scriptsize Tableau Research}

\teaser{
  \centering
  \includegraphics[width=0.99\linewidth]{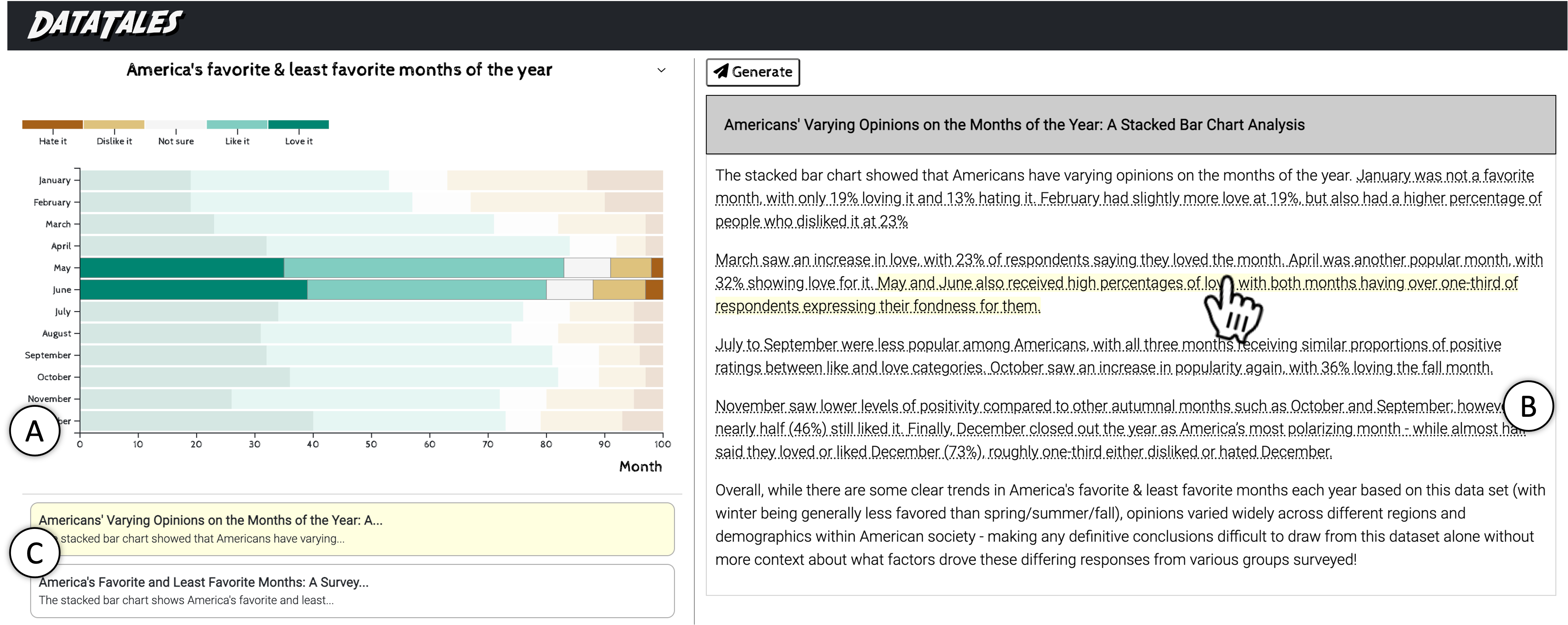}
  \vspace{-1em}
  \caption{\datatales~user interface including an interactive visualization (A), generated stories panel (B), and a history of generated stories (C).
  Here, as the user hovers the cursor over a sentence in the generated story, the system dynamically highlights the marks corresponding to the text (in this case, the bars for the months of May and June).
}
  \label{fig:interface}
}

\abstract{
Authoring data-driven articles is a complex process requiring authors to not only analyze data for insights but also craft a cohesive narrative that effectively communicates the insights.
Text generation capabilities of contemporary large language models (LLMs) present an opportunity to assist the authoring of data-driven articles and expedite the writing process.
In this work, we investigate the feasibility and perceived value of leveraging LLMs to support authors of data-driven articles.
We designed a prototype system, \datatales, that leverages a LLM to generate textual narratives accompanying a given chart.
Using \datatales~as a design probe, we conducted a qualitative study with 11 professionals to evaluate the concept, from which we distilled affordances and opportunities to further integrate LLMs as valuable data-driven article authoring assistants.
} 

\CCScatlist{
  \CCScatTwelve{Human-centered computing}{Visu\-al\-iza\-tion}{Visualization design and evaluation methods}{}
}

\usepackage{soul}
\usepackage{fontawesome5}

\newcommand{\datatales}{\textsc{DataTales}}

\newcommand{\cre}[1]{\textcolor{black}{#1}}

\begin{document}


\firstsection{Introduction}

\maketitle



\textit{Data-driven articles} that feature primarily textual narratives containing claims and insights backed by data and illustrated with data visualizations are a popular means of communication in fields like journalism and business reporting~\cite{sultanum2021leveraging}.
Authoring data-driven articles, however, is often a complex and tedious process.
Authors need to analyze the data to identify insights, order insights in an appropriate sequence, and write a cohesive narrative to communicate those insights with effective transitions and appropriate domain context.

The emergence of contemporary large language models (LLMs) and their remarkable text generation capabilities led to increased interest in assessing their value for a range of creative writing tasks~\cite{fang2023systematic}, including data storytelling~\cite{li2023ai}. 
While this technology has the potential to fundamentally reshape the way people use writing tools~\cite{spencer2023future},
it also introduces news challenges such as unreliable outcomes, lack of domain understanding, \cre{prompt complexity}, ethical concerns, among others~\cite{li2023ai}. We believe that these issues require thoughtful design solutions to circumvent them, and that different writing genres may benefit from purpose-specific features built around these models. 

In this work, we investigate the potential of LLMs to support the authoring of data-driven articles. Based on the deep intertwining of charts and text in these articles, \cre{and targeting the intermediate stages of the visual storytelling process where authors are actively building a story based on exploratory findings~\cite{lee2015more},} we propose \textit{chart interaction} as a more intuitive alternative to direct prompting for conveying narrative intent to the LLM. We developed \cre{an early proof-of-concept}, \datatales, that generates textual content for an accompanying chart, and additionally allows authors add \textit{chart annotations} to guide focus of the story. Authors can use the generated text as-is, edit portions of the text, or generate multiple instances to pick-and-choose what they like.
Using \datatales~as a design probe, we conducted a qualitative study with 11 data professionals to understand the benefits and challenges of this concept to support authoring of data-driven articles. The concept was well received even in this preliminary form, showcasing potential for further investigations. We discuss lessons learned and directions for future development in the form of takeaways to share with the community.
\section{Related Work}

\noindent\textbf{Authoring data-driven articles.}
There is a significant body of research on narrative visualization and storytelling~\cite{segel2010narrative,stolper2016emerging,lee2015more}, a subset of which has specifically focused on authoring data-driven articles.
For instance, one line of work has explored the use of markup language-based frameworks to support the authoring of interactive articles on the web~\cre{\cite{conlen2021idyll,latif2018exploring,latif2019authoring}}.
Another set of systems like Kori~\cite{latif2021kori}, VizFlow~\cite{sultanum2021leveraging} and DataParticles~\cite{cao2023dataparticles} adopt a more graphical and mixed-initiative approach and allow authors to configure interactive links between text and charts while authoring data articles.
Besides systems that explicitly focus on content drafting and presentation fine-tuning, another body of work also includes data fact- or insight-recommendation systems that suggest singleton takeaway statements for visualizations during the data exploration phase to help authors identify talking points in their articles~\cite{srinivasan2018augmenting,shi2020calliope,wang2019datashot, li2023notable}.
Our work furthers the line of research on authoring data-driven articles by investigating the use of contemporary LLMs to generate ideas for textual content that authors can further edit.



\noindent\textbf{LLMs in data visualization and writing.}
Recent advances in model architectures, performance, and availability have led to a surge of LLM-based applications for writing support. These include creative writing support tools such as plot suggestions~\cite{singh2022hide}, journalistic angle ideation~\cite{petridis2023anglekindling} and co-writing of theater scripts~\cite{mirowski2023co}; as well as technical writing support such as 
argumentative writing~\cite{zhang2023visar}, scientific writing~\cite{gero2022sparks}, and reverse outlining for manuscript revision~\cite{dang2022beyond}.
Within the data visualization space, LLMs have been used to power natural language interfaces~\cite{shen2021towards} for 
visualization authoring~\cite{wang2022towards}.
Only a few works have looked at text content generation in a data visualization context, to create data stories from a set of user-provided keyframe data facts~\cite{sun2022erato}, and natural language summaries of a given chart for accessibility purposes~\cite{obeid2020chart}.
To our knowledge, our work is the first to look specifically at leveraging LLMs for data-driven articles. 

\section{\datatales}


\begin{figure}
    \centering
    \includegraphics[width=0.85\linewidth]{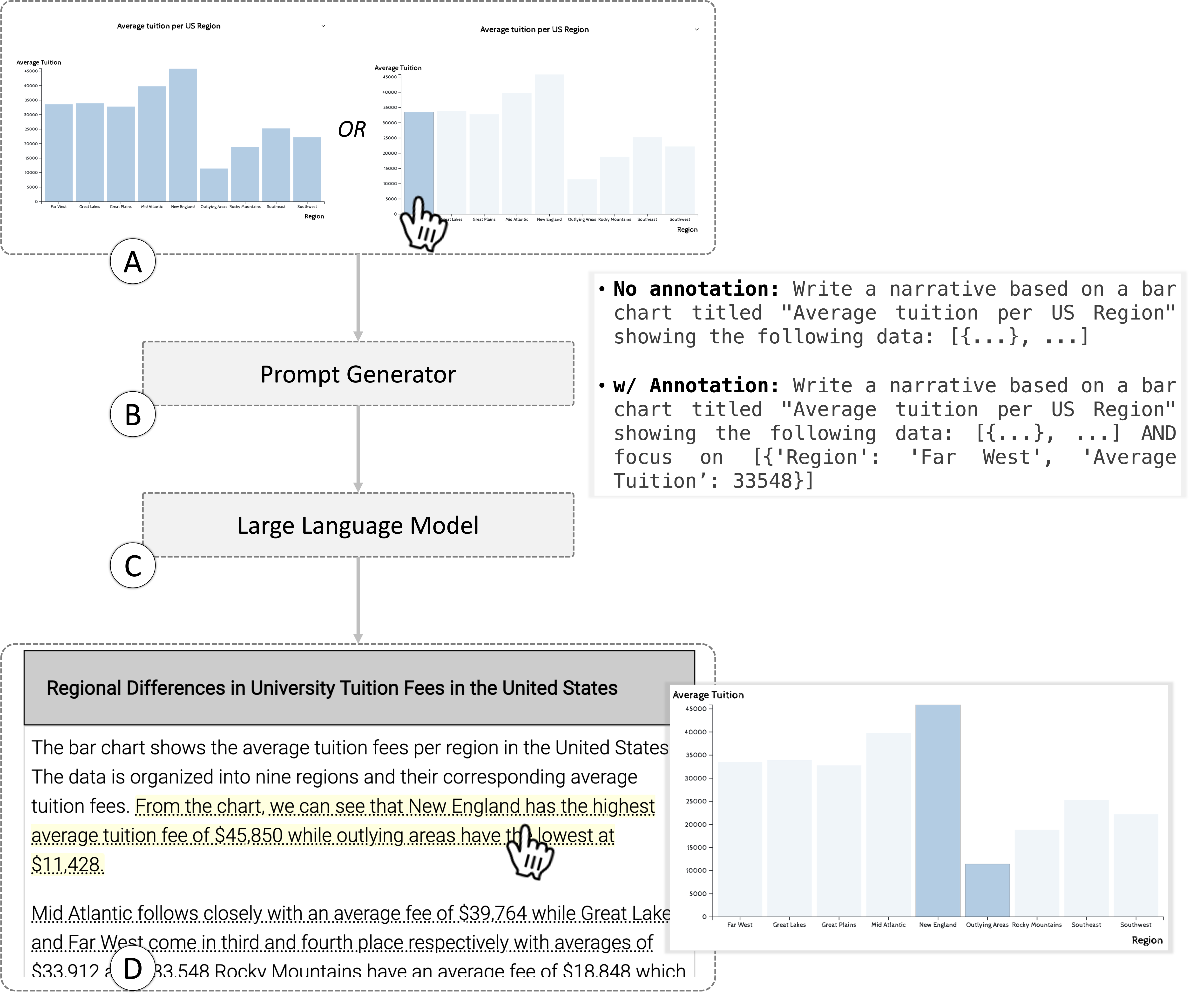}
    \vspace{-1em}
    \caption{\datatales~workflow overview. Given a chart and an optional set of annotations, the system generates textual narratives that are interactively linked to the chart and can be further edited by authors.}
    \label{fig:workflow}
    \vspace{-1em}
\end{figure}

Fig.~\ref{fig:interface} shows the \datatales~user interface and Fig.~\ref{fig:workflow} summarizes the system workflow.
The system is implemented as a web application using a React and  Python Flask setup. \cre{It features a curated list of datasets, with respective charts} rendered using D3.js. 
For the language model, we use the OpenAI API for the `\texttt{gpt-3.5-turbo}' model
~\footnote{The latest model available for development at the time of this research.}.
Below we detail workflow steps and core system features.

\vspace{.5em}
\noindent\textbf{Chart and user annotations.}
\datatales~covers a wide array of charts commonly found in data-driven reports and articles, including bar charts with variants like stacked and group bars, scatterplots, single- and multi-series line charts, and choropleth maps.
The current implementation contains a set of predefined charts covering a breadth of datasets including demographic survey responses, unemployment rates, automobile data, and Olympic medal winner history, among others.
When asking the system to generate a story (via the \texttt{Generate} button, Fig.~\ref{fig:interface}B), authors can have the entire chart considered for input or optionally add annotations to guide the \cre{LLM} to emphasize specific data points or ranges when generating its response.
For instance, Fig.~\ref{fig:workflow}A shows an example where an author highlights two bars they want the system to focus on when generating a story.
\datatales~supports various annotations including mark selection, color legend range selection, and axis range selection, \cre{which can be combined for more complex guidance}~\cite{ren2017chartaccent}.

\vspace{.5em}
\noindent\textbf{Prompt generation.}
The key idea underpinning \datatales~is that a system can take a chart or an annotated chart as input and leverage a LLM to recommend data-driven narratives.
To this end, we iterated on several template variations to generate the prompts that are fed into the LLM.
Specifically, we explored different features to include (or exclude) in the prompt such as the chart type, encodings, analytic tasks associated with specific charts~\cite{saket2018task,kim2018assessing,zacks1999bars}, the chart title, the underlying dataset metadata, and user annotations, story length, among others.
Besides experimenting with features, we also tried different phrasings to \cre{assess} if the order of features or the grammar of the prompt notably impacted the generated narrative.

We generated 10-20 narratives for each chart type with different combinations of these features.
Inspecting the results, we iteratively excluded or combined features that yielded redundant results.
For instance, we noticed that including the encoding information in the prompt generated statements reiterating the chart.
We thus excluded encoding details from the prompt as such statements tend to offer little value to readers~\cite{lundgard2021accessible}.
Similarly, we initially experimented with including analytic tasks (e.g., finding extremes, identifying correlations).
However, we noticed that including the chart type in the prompt (e.g., `bar chart', `scatterplot', `line chart') resulted in narratives comparable to those generated by including analytic tasks.
Correspondingly, keeping in mind the simplicity and brevity of specifying the chart type (over analytic tasks), we only included that information in the final prompt template.
Fig.~\ref{fig:workflow}B shows an example of the prompts generated by \datatales~but the general template for generating data narratives is as follows:
\vspace{-0.3em}
\begin{quote}
    \small
    \texttt{Write a narrative based on a [chartType] showing the following data: [chartData] on the topic "[chartTitle]" focusing on: [chartAnnotations*]}
\end{quote}
\vspace{-0.3em}
where, \texttt{*} indicates an optional parameter that is included in the prompt only if it is available in the input chart.
\texttt{chartData} is the data array that is bound to the marks and \texttt{chartAnnotations} is a list of data items for selection annotations (e.g., $\{Year:2000, Country:Australia\}$) and/or values in the case of axis brush annotations (e.g., $\{Year~between~[1980, 2001]\}$).
Once a narrative is generated, we prompt the LLM again to generate a title: 
\vspace{-0.3em}
\begin{quote}
    \small
    \texttt{Suggest a title for the following narrative: [narrativeText]}.
\end{quote}
\vspace{-0.3em}
The title and text are sent as to the system front-end as a self-contained story.
\cre{These prompts generated reasonable results for our purposes, although we argue that further experimentation with prompt patterns~\cite{white2023prompt} would be worthwhile.}   

\begin{figure}[t!]
    \centering
    \includegraphics[width=.9\linewidth]{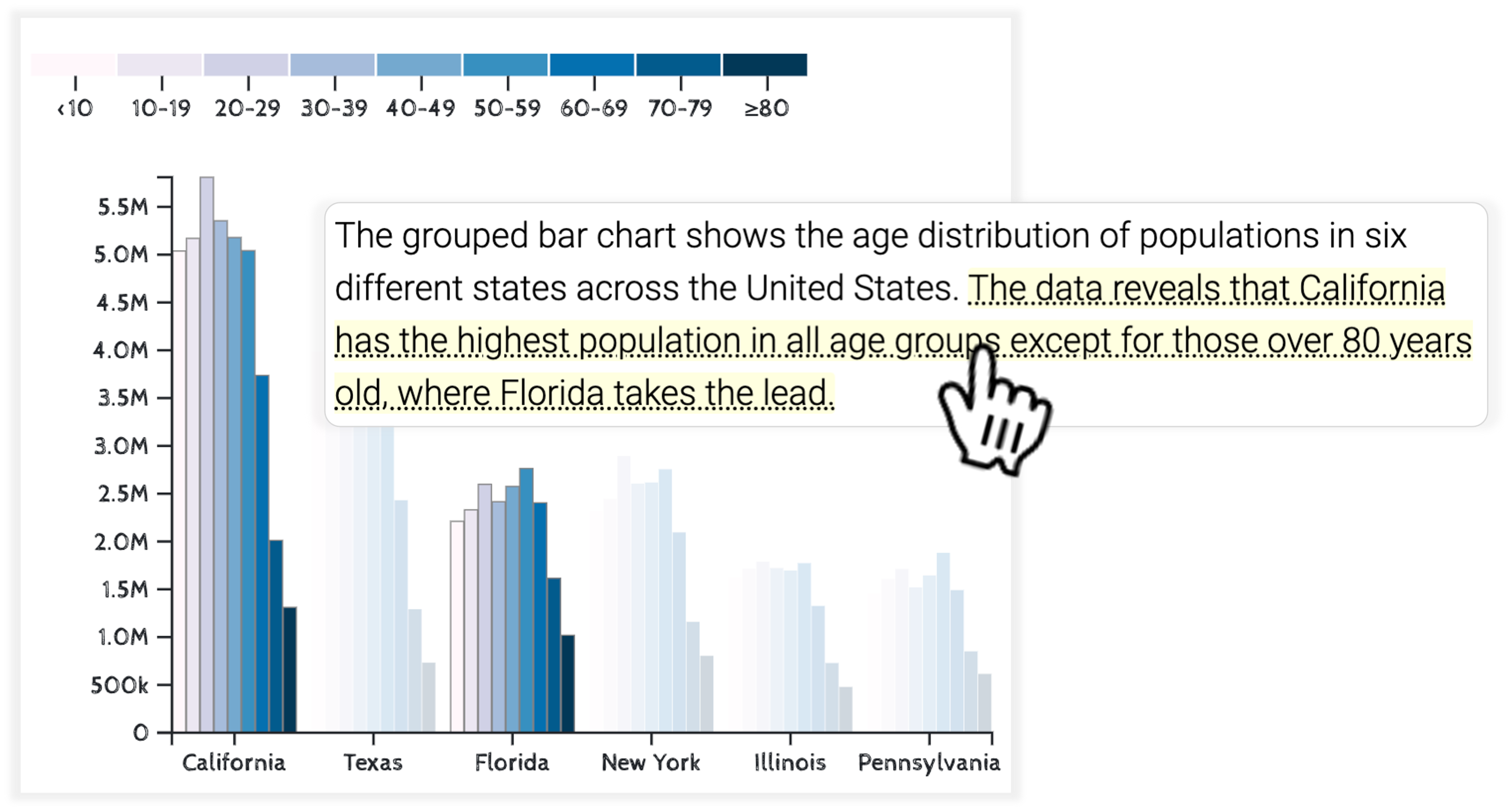}
    \vspace{-1em}
    \caption{Example of an incorrect statement generated by the LLM (contrary to the text, the chart shows that Florida \textit{does not} have a higher number of people over the age of 80 compared to California).
    The text$\rightarrow$chart linking feature helps verify the statement and identify the erroneous interpretation by dynamically highlighting the two states.}
    \label{fig:eg-incorrect}
    \vspace{-1em}
\end{figure}

\vspace{.5em}
\noindent\textbf{Linking the generated text to the input chart.}
Once the LLM generates the narrative, \datatales~proactively processes the generated story to identify data references.
\cre{Similar to prior natural language systems for visualization (e.g.,~\cite{gao2015datatone,narechania2020nl4dv}), we use a combination of dependency parsing and keyword matching to map phrases in a sentence to attributes and values in the visualized data.}
\datatales~highlights whole sentences containing data references using a dotted underline to emphasize that the sentence talks about a specific set of marks on the chart.
To aid reading and comprehension, and incorporating ideas from prior work on interactively linking text and charts~\cite{latif2021kori,srinivasan2018augmenting,kong2014extracting,sultanum2021leveraging}, as authors hover on these underlined sentences, \datatales~highlights relevant portions of the chart (see Figs.~\ref{fig:interface} and~\ref{fig:workflow}D).
Besides improving readability, our motivation to include this text$\rightarrow$chart linking was also that visually seeing the data being referred to in the text could serve as a quick verification for potential hallucinations or incorrect interpretations by the LLM (e.g., Fig.~\ref{fig:eg-incorrect}).
Authors can then redact the stories themselves, and their edits are shown in a different italicized format. 

\section{Evaluation}

To assess whether our envisioned concept of using chart  interaction for LLM story generation made sense in the context of data-driven articles, we conducted a qualitative user study. We used \datatales{} as a \textit{design probe} to expose participants to this concept in the context of an authoring task, and then elicited feedback on the experience. 

We recruited 11 data professionals (P1-P11) with prior experience in authoring data-driven articles or similar reports. Backgrounds encompassed content writers, dashboard designers, project managers and consultants, spanning multiple organizations. Participants were recruited via slack communities on related interest channels. Incidentally, most participants had some prior exposure to LLM-based tools, and five reported using these tools for authoring support at some point (e.g., for brainstorming, starting points, outlines,~summaries).

Feedback sessions entailed a 20-min data story authoring task on a given chart \cre{followed by a semi-structured interview to discuss their experiences with the tool. In lieu of conducting data analysis from scratch, participants were given one of a subset of four distinct \cre{dataset}+chart types available in the tool (including stacked bar chart, line chart, scatterplot, and choropleth), which helped standardize experiences and keep sessions concise.} Participants were told to use \datatales{} in their authoring process however they saw fit, while editing their working draft in a separate document editor
for maximum editing flexibility. We also encouraged them to think aloud whenever possible during the authoring task.

This study setup provided us with rich qualitative data to assess potential and limitations of LLMs for this task. We organize our findings in the form of \textbf{takeaways (T1-T13)}, encompassing observations of authoring workflows (Section~\ref{sec:authoring-workflows}), perceived value and affordances of LLM-based article authoring systems like \datatales{} for \cre{data-driven} authoring experiences (Section~\ref{sec:affordances}), and identified limitations plus potential solutions (Section~\ref{sec:opportunities}).

\subsection{Authoring Workflows}
\label{sec:authoring-workflows}
Lessons and ideas emerging from task observations are as follows.

 \textbf{(T1) A master draft + multiple stories.} Participants were free to decide a format and framing for their stories, which led to a diverse set of data-driven articles. That said, authoring workflows were fairly consistent across participants: \cre{they all} generated multiple versions, reused chunks from one or more versions\,---\,occasionally a whole story, but most frequently short paragraphs from various ones\,---\,rearranged them, and then made editorial revisions for style and flow.  This suggests a setup for managing multiple story generation outcomes, and \cre{that} maintaining an integrated master draft \cre{which} can be easily populated with generated segments would be useful. 

\textbf{(T2) Expediting error checks.} As expected of LLM output, a number of inaccuracies~\cite{borji2023categorical} were spotted, prompting them to carefully check generated text for errors: \textit{``is this legit?"} (P3). The text-chart highlights were frequently used for this purpose and several agreed on its usefulness (P3, P8, P9), suggesting text-chart readability aids~\cite{latif2021kori}  should be further explored. On that note, some folks appreciated how author changes were explicitly signaled (P2, P5), helping retain context of what was fixed and what needs checking.

\textbf{(T3) Synergies between chart and text.} 
Participants extensively leveraged chart interaction for their story generation: from an average of 4 stories per person, about 3 featured annotations. While the annotated stories did not always feature the depth and framing participants were hoping for (more under \textbf{T9}, \textbf{T10}), it consistently matched the selections (and author intent, as per think-aloud feedback), and results were still often usable and repurposed. Some participants also explicitly acknowledged the value of having the chart integrated into their drafting environment (P6, P7). These findings suggest that \datatales{}'s use of chart interaction for text generation shows promise and is worth exploring further.

\textbf{(T4) Coupling of annotations and generated stories.} Annotations were frequently used to get more details on selected data features,  usage that was largely intuitive and represented important context for the text. We posit that preserving annotation context for generated snippets on a master draft would be very useful, not only for authors to recall provenance of snippets but also to potentially reuse as embedded highlights for readers, e.g., in the context of a dynamic story format such as scrollytelling~\cite{sultanum2021leveraging}.

\textbf{(T5) Potentially time saving.} While participants were not expected to finish their stories within the 20min, 3 of them successfully completed a first draft in the allotted time, suggesting potential efficiency gains in the authoring process. Several participants could also foresee saving time in the long run, e.g., \textit{``would cut out a good 15 to 20 minutes of my work''} (P6), and getting a head start on the writing, e.g., \textit{``Getting started is sometimes hardest thing (...). I'll be looking at the data, procrastinating, trying to find correlations and relationships(...). And it does that for me, at least a base level''} (P8).



\subsection{Affordances}
\label{sec:affordances}

Despite the limited nature of the tool as a proof-of-concept design probe, participant reactions to the experience ranged from congenial to enthusiastic. Rationales on how \datatales{} supported their authoring experience in new and positive ways are compiled below.

\textbf{(T6) Insights over data facts.} While data facts are an important part of a data story, the segments most often repurposed and appreciated by participants were those containing level-3 and level-4 statements in Lundgard and Satyanarayan's categorization of chart descriptions~\cite{lundgard2021accessible}, which participants referred to as ``\textit{the why's}'' (P3, P10, P11). For example, on a dataset about cars acceleration vs. horsepower vs. country of origin, this could include things like identifying trends (e.g., \textit{``cars with higher horsepower tend to have better acceleration rates'}'), conclusions following findings (e.g., \textit{``The US auto market prioritizes higher horsepower''}), and external context (e.g., \textit{``policymakers should consider regulating emissions for consumers who value speed over efficiency''}). Several added that aggregating this \textit{``human knowledge''} was one of the most valuable aspects of the experience, complementing their authoring work with new information (P1, P3), alternative framings (P3, P7), and confirmation of current viewpoints (P3, P11).

\textbf{(T7) Explanatory support: what to talk about and how}. 
Getting a first initial draft is challenging, and having a range of full stories available helped provide starting points to overcoming writer's block (P6, P8, P10, P11): \textit{``I didn't know where I wanted to approach it, and then after generating a couple stories, I saw a trend and decided that'd be my focus"} (P6).  New ideas or context present in those stories also provided inspiration for new directions to explore (P5, P7), e.g., the extent that \textit{``the Great Depression''} affected US gold medal performance in the Summer Olympics (P7); as well as seeing familiar findings or terms portrayed in a different ways, with evocative phrasings, e.g., \textit{``powerful muscle cars"} (P11), and unique ordering of findings (P3). We argue there may be value in not only supporting easy generation and management of different stories, but also allowing for more diversity across different stories (e.g., via a slider to control the model's \texttt{temperature}), even at the cost of more spurious findings.


\textbf{(T8) Exploratory support: a different lens on the data}. 
An unexpected use of \datatales{} was as a data exploration tool, e.g., to form hypothesis (P3), to gather facts (P7) and to get a high-level summary of the data in natural language form (P2, P6, P9). While our study setup induced some analysis as participants were asked to work with an unknown \cre{dataset}, several of the dynamics observed applied to in-between stages of analysis and storytelling: e.g., getting ideas for additional datasets and facts to look into (P7, P8), and using annotations to dig deeper into individual data points (P11). Several participants also leveraged \datatales{} to ``test'' hypotheses by confirming or denying prior assumptions (P3, P7, P10), and to \textit{``ask its opinion''} (P7, P11).  
This showcases the intertwined nature of analysis and storytelling, and how authoring tasks can benefit from interactive visualizations integrated into the drafting environment.

\subsection{Opportunities}
\label{sec:opportunities}
While overall reactions to the tool were net-positive, participants also raised several concerns and suggestions for improvement, informing many compelling directions for future work. 

\textbf{(T9) More control over overall style.} A prominent pain point was the lack of control over \textit{voice} (e.g., corporate voice (P5), a business owner's perspective (P3)), \textit{tone} (e.g., formal vs. personal (P2), make it less ``robotic'' (P8)), and \textit{format} (e.g., organize findings from highest to lowest counts (P4)). Generated stories were often found \textit{``too wordy''} (P5-P7, P10, P11), requiring heavy editing to cut them down. Those with prior LLM exposure suggested bridging this gap by writing or editing underlying prompts generated by the tool (P5, P6, P10); on the other hand, it was remarked that prompting could be found too intimidating or unfamiliar for other authors (P3), which calls for some form of reasonable middle ground. We envision \datatales{} could provide predefined fields for authors to describe target audience, voice, and format in natural language, which could then be incorporated into the base template to generate a new story.

\textbf{(T10) Co-writing micro-tasks.} Apart from overall style, participants also wanted assistance in generating paragraph-level content for targeted insights  (P4, P5, P6), e.g., \textit{``let's focus on January thru March, and talk about the holiday angle"} (P5) and \textit{``tell me why the US has such an outlier number of cars on the higher horsepower end"} (P6). While power-users could again benefit from direct prompting here, P5 suggested a more seamless  \textit{fill-in-the-blanks} approach, designating spaces for automatic completion for diegetic prompting~\cite{dang2023choice}, i.e., leveraging parts of the writing itself. For data-centered insights, we envision combining chart interaction with natural language input, e.g., by editing the chart title to nudge generation of different insights, drawing trend lines over the chart to indicate what patterns are important to focus on, and adding descriptive annotations to particular data points to retrieve targeted context. 

\textbf{(T11) More exploratory aids.} Many participants wanted to take their insights a step further by getting recommendations for external related datasets to include in the story (P3-P7, P11),  and potential comparisons to be drawn within or across datasets (P2, P4-P7, P11). Some also framed these recommendations as an opportunity to continue learning about the data \textbf{(T8)} and helping them formulate \textit{``follow up questions"} (P3) and \textit{``additional areas of exploration''} (P5), giving authors more directions to consider for their stories.

\textbf{(T12) Summarize stories.} Another consequence of ``wordy'' stories \textbf{(T9)} is they take \textit{``a lot of time to read through''}~(P5), which is specially onerous when trying to generate stories for exploratory analysis purposes \textbf{(T8)}, i.e., natural language overviews of the data. Some participants suggested an option to depict stories in a more concise format (P2, P5, P7, P11), e.g., bullet-point form (P2, P5). Beyond data analysis overview, these summaries could also allow for quicker inspection of new stories,  faster recall of past stories, and support revisions of a master draft via reverse outlining~\cite{dang2022beyond, king2012reverse}. 

\textbf{(T13) Internal knowledge, external validation.}
As stated earlier, a number of inaccuracies were surfaced in the generated stories \textbf{(T2)}. These include objective errors (e.g., such as hallucinations, failure to spot obvious patterns, labeling mix-ups), but also subtle mistakes that are hard to spot without expert background: e.g., on a misguided reasoning for Vermont's low unemployment rates, P8 warned \textit{``if I didn't have local (Vermont) knowledge, I'd just have taken this at face value"}. To mitigate this, automatic retrieval of supporting references was suggested, e.g., source citations (P9) and external datasets for triangulation (P8, P9). Some participants also wanted to provide corrective feedback to the model (P3, P6), on the assumption that it would get better \textit{``the more you use it"} (P6).
Another related concern was trying to sort out what statements are derived from the underlying chart data and what was brought in by the LLM as external context (P2, P5, P9). Similar to how manual edits are clearly marked \textbf{(T2)}, participants suggested further signaling passages with a different color or formatting to denote source.


\section{Reflections}


Following our takeaways \textbf{(T1-T13)}, we conclude with a few thoughts that emerged from discussions \cre{that} are pertinent to the use of LLMs for authoring and visual storytelling more generally.  



First, we consider \textbf{bias versus framing.} One of the challenges reported by participants was how difficult it was at times to get generated text that matched the points they wanted to make in a story, i.e., \textit{``confirm their viewpoints''}. On one hand, \textit{``picking a side''} is an integral exercise of telling a story. On the other, we also know how prone LLMs are to fabricating evidence in a convincing manner. As such, to honor the overall purpose of data stories to stay true to the data (within the many \cre{valid} interpretations that may arise), and given the significant exploratory component that the authoring process \cre{entailed}, it is essential we consider ways to \cre{mitigate} confirmation bias and amplification of harmful speech. 

A related point is the importance of \textbf{bridging data literacy}. It was raised in participant discussions how folks with lower data literacy skills would benefit the most from authoring support (P5, P9), but at the same time, having these skills would be crucial to properly guard against LLM errors (P8). 
Under the same tenet of mitigating misinformation, we should consider ways to better equip authors with critical thinking skills: from facilitating fact checking and encouraging all statements to be fact-checked, to nudging techniques that encourage reflection, e.g., phrasings that ``ask a question'' instead of stating ``what and why''~\cite{danry2023don}.


And finally, we make a case for the use of \textbf{LLMs alongside classic techniques.} While our studies demonstrated how LLMs can add value to the authoring by easily incorporating context and insights, their unpredictability does offset some of the efficiency it is purported to bring. Many of the solutions suggested to \cre{deal with} this uncertainty would call for add-on modules to parse the generated text, identify entities, link text entities and chart elements, retrieve external evidence, and so on. While there is ongoing work in applying LLMs to natural language tasks like information extraction~\cite{li2023evaluating} and fact-checking~\cite{zhang2023interpretable}, we argue that tried and tested heuristics-based algorithms and classic NLP techniques will still play an essential role in the foreseeable future.






\bibliographystyle{abbrv-doi}

\bibliography{references}
\end{document}